\documentstyle[aps,prb,epsfig]{revtex}
\begin{document}
\baselineskip=0.5cm
\renewcommand{\thefigure}{\arabic{figure}}
\title{Single-particle density matrix and superfluidity in the two-dimensional
Bose Coulomb fluid}
\author{A. Minguzzi$^1$, B. Davoudi$^{1,2}$ and M. P. Tosi$^1$}
\address{
$^1$NEST-INFM and Classe di Scienze, Scuola Normale Superiore, I-56126 Pisa, Italy\\ 
$^2$Institute for Studies in Theoretical Physics and Mathematics, Tehran, P.O.Box 19395-5531,Iran\\
}
\maketitle
\begin{abstract}
A study by W. R. Magro and D. M. Ceperley [Phys. Rev. Lett. {\bf 73}, 826 (1994)] has shown that the ground state of the two-dimensional fluid of charged bosons with logarithmic interactions is not Bose-condensed, but exhibits algebraic off-diagonal order in the single-particle density matrix $\rho(r)$. We use a hydrodynamic Hamiltonian expressed in terms of density and phase operators, in combination with an $f$-sum rule on the superfluid fraction, to reproduce these results and to extend the evaluation of the density matrix to finite temperature $T$. This approach allows us to treat the liquid as a superfluid in the absence of a condensate. We find that (i) the off-diagonal order arises from the correlations between phase fluctuations; and (ii) the exponent in the power-law decay of $\rho(r)$ is determined by the superfluid density $n_s(T)$. We also find that the plasmon gap in the single-particle energy spectrum at long wavelengths decreases with increasing $T$ and closes at the critical temperature for the onset of superfluidity.
\end{abstract}

\vspace{1cm}

\section{INTRODUCTION}
The fluid of point charges interacting via a logarithmic Coulomb potential in a strictly two-dimensional (2D) space is a basic model in statistical mechanics with main relevance to the theory of vortex fluctuations in superfluid or superconducting films\cite{1}. Magro and Ceperley\cite{2} have discussed the bosonic ground state of this model system. They first used a sum-rule argument from the work of Pitaevskii and Stringari\cite{3} to show that the zero-point fluctuations associated with long-wavelength plasmons rule out the presence of a Bose-Einstein condensate even at zero temperature. They proceeded to evaluate the single-particle density matrix $\rho(r)$ by quantum Monte Carlo (QMC) methods and showed that the ground state exhibits algebraic off-diagonal long-range order: the decay of $\rho(r)$ with increasing distance $r$ is through the power law $r^{-r_s/4}$, where $r_s$ is a coupling strength parameter determined by the areal density $n$.

	As Magro and Ceperley\cite{2} emphasize, although the ground state of the 2D Bose Coulomb fluid (2D-BCF) is not condensed, superfluidity is nonetheless possible. If the model is indeed superfluid, it may then provide an ideal system in which to study the differences between superfluidity and Bose-Einstein condensation.
	
	A number of structural and collective dynamical properties have been calculated for the 2D-BCF within approximate theoretical schemes\cite{4}, but no other studies seem to have been made of its momentum distribution and one-body density matrix beyond that of Magro and Ceperley. In the present paper we evaluate these properties for the charged fluid in the weak-coupling regime corresponding to high density, by adopting a hydrodynamic reduction of its Hamiltonian as previously proposed by Popov\cite{5} for 2D neutral Bose fluids (see also Meng\cite{6}). A crucial point of our approach is to allow for a difference between the long-wavelength dispersion relations of single-particle and collective excitations at finite temperature $T$, through the use of distinct $f$-sum rules on the superfluid and the total particle number densities\cite{7}. We recover by this approach the results of Magro and Ceperley\cite{2} for the power-law decay of $\rho(r)$ in the ground state and obtain their extension to finite temperature. We show that within this approach the 2D-BCF is indeed superfluid at $T = 0$ and that in the weak-coupling regime a slowly declining superfluid fraction persists as the temperature is raised. In fact, the superfluid fraction enters to determine the value of the exponent for the power-law decay of $\rho(r)$. The presence of the superfluid density in the power-law decay of the density matrix in a neutral 2D Bose gas was already argued by Popov\cite{5}.
	
	Our present progress in understanding the properties of the 2D-BCF borrows from the theory of a quasicondensate state in the neutral 2D Bose gas, that was developed by a number of authors\cite{5,8,9,10,10a} and has also been used within a Bogoliubov approach to describe the 2D fluid of charged bosons with $e^2/r$ interactions\cite{11}. In essence, the local properties of a quasicondensate, over distances that are shorter than the phase correlation length, are the same as those of a genuine condensate. However, in our treatment of the 2D-BCF we dispense with the notion of a quasicondensate fraction and base our arguments on the $f$-sum rule for the superfluid fraction.
	
	The paper is organized as follows. In Section II, after introducing the Hamiltonian of the charged fluid with its neutralizing background and recalling the $f$-sum rules for a superfluid from the work of Hohenberg and Martin\cite{7}, we present the hydrodynamic Hamiltonian as expressed in terms of density and phase operators. Following Meng\cite{6} we diagonalize it in the momentum representation and obtain a dispersion relation for single-particle excitations which has the Bogoliubov form but contains the superfluid density in place of the condensate density. In Section III we use the hydrodynamic Hamiltonian to determine the power-law decay of the one-body density matrix from the behavior of the correlations between phase fluctuations entering the single-particle Green's function. In Section IV we evaluate $\rho(r)$ by using the resummation method adopted by other authors\cite{8,9} to deal with a fluid with a quasicondensate, and show that the power-law decay obtained from the single-particle Green's function is recovered. We then report in Section V numerical results for the superfluid density from Landau's formula and for the one-body density matrix as functions of temperature at various values of the coupling strength $r_s$ in the weak-coupling regime. Section VI concludes the paper with a brief summary. In an Appendix we show that the standard Bogoliubov approach\cite{12} would indeed yield that the condensate fraction is zero in the ground state of the 2D-BCF.
	
\section{HYDRODYNAMIC HAMILTONIAN AND SINGLE-PARTICLE EXCITATIONS}
The 2D-BCF is described by the Hamiltonian
\begin{equation}
H=\int d{\bf r}\;\psi^\dagger({\bf r})\left [-\frac{\nabla^2}{2m}-\mu\right ]\psi({\bf r})
+\frac{1}{2}\int d{\bf r}\int d{\bf r}'\psi^\dagger({\bf r})\psi^\dagger({\bf r}')V(|{\bf r}-{\bf r}'|)
\psi({\bf r}')\psi({\bf r})
\end{equation}
	(having set $\hbar = 1$), where $\psi(r)$ is the field operator, $\mu$ is the chemical potential and $V(r)$ is the interaction potential. This is the solution of the 2D Poisson equation $\nabla^2V(r)=-2\pi e^2 \delta({\bf r})$, yielding $V(r) = -e^2\ln(r/l_0)$ where $l_0$ is a reference length that we shall take as $l_0 =(me^2)^{-1/2}$. The Fourier transform of the potential is $V_{\bf k}=2\pi e^2/k^2$, and in the following we shall set $V_{{\bf k}=0}= 0$  on account of a uniform background ensuring global charge neutrality\cite{13}. The coupling strength is measured by the dimensionless parameter $r_s$, which is defined by $r_s=(2me^2/\pi n)^{1/2}$.
	
	A special role will be played in the following by a sum rule involving the superfluid density $n_s$. We first recall that the usual $f$-sum rule\cite{14} involving the particle density $n$ can be recast through the continuity equation into a sum rule on the longitudinal current-current response function $\chi_{JJ}({\bf k},\omega)$,
\begin{equation}
\int_{-\infty}^\infty\frac{d\omega}{\pi}\frac{{\rm Im} \chi_{JJ}({\bf k},\omega)}{\omega}=\frac{n}{m}.
\end{equation}
For a charged fluid this is equivalent to the well-known plasmon sum rule on the longitudinal electrical current density. In addition to Eq.~(2), a sum rule involving the response function of the superfluid velocity $v_s$ holds in a superfluid\cite{7,14}, which reads
\begin{equation}
\lim_{{\bf k}\rightarrow 0}\int_{-\infty}^\infty\frac{d\omega}{\pi}\frac{{\rm Im} \chi_{v_sv_s}({\bf k},\omega)}{\omega}=\frac{1}{mn_s}.
\end{equation}
The role of this relation as a long-wavelength $f$-sum rule on the superfluid response becomes evident from its similarity to Eq.~(2) when it is rewritten as a sum rule for the superfluid particle current density ${\bf J}_s=n_s{\bf v}_s$.

	We are now ready to proceed to a hydrodynamic reduction of the Hamiltonian (1), following the work of Popov\cite{5} and Meng\cite{6} on neutral superfluids. This is obtained by making the transformation
\begin{equation}
\psi({\bf r})=\sqrt{\hat{\rho}({\bf r})}\exp{[i\Phi({\bf r})]},\;\;\;\;\;\psi^\dagger({\bf r})=\exp{[-i\Phi({\bf r})]}\sqrt{\hat{\rho}({\bf r})}
\end{equation}
for the components of the field operators which correspond to wave number $k$ below a cut-off $k_o$\cite{5}. Here $\hat{\rho}({\bf r})$ and $\Phi({\bf r})$ are the particle density and phase operators. The transformed Hamiltonian can be brought to a quadratic form by considering only small fluctuations in $\hat{\rho}({\bf r})$ around a constant value $\rho_0$,
\begin{equation}
\hat{\rho}({\bf r})=\rho_0+\eta({\bf r}).
\end{equation}
A term linear in $\eta({\bf r})$ can be dropped by setting $\mu = V_{{\bf k}=0}=0$, and we obtain the hydrodynamic Hamiltonian
\begin{equation}
H_h=\int d{\bf r}\left [(8mn_s)^{-1}\left (\nabla\eta({\bf r}) \right)^2+(n_s/2m)\left (\nabla\Phi({\bf r}) \right)^2\right]
+\frac{1}{2}\int d{\bf r}\int d{\bf r}'V(|{\bf r}-{\bf r}'|)\eta({\bf r})\eta({\bf r}').
\end{equation}
Here, the constant $\rho_0$ has been taken equal to $n_s$ in order to satisfy the sum rule (3), as we shall explicitly demonstrate at the end of this Section. As usual, the superfluid velocity is ${\bf v}_s=\nabla\Phi({\bf r})/m$. We stress that the Hamiltonian in Eq.~(6) describes only long-wavelength fluctuations of small amplitude in the superfluid density and velocity field.

	The Hamiltonian (6) can be diagonalized by first expanding the density fluctuation and phase operators in the form
\begin{equation}
\eta({\bf r})=\sum_{{\bf k}(k<k_0)}q_{\bf k} \exp{(i{\bf k}\cdot{\bf r})},\;\;\;\;\Phi({\bf r})=\sum_{{\bf k}(k< k_0)}p_{-{\bf k}} \exp{(i{\bf k}\cdot{\bf r})},
\end{equation}
which yields
\begin{equation}
H_h=\sum_{{\bf k}(k<k_0)}\left\{\left[(k^2/8mn_s)+\frac{1}{2}V_{\bf k}\right]q_{\bf k}q_{-{\bf k}}+(n_sk^2/2m)p_{\bf k}p_{-{\bf k}}\right\}.
\end{equation}
We next set $q_{\bf k}=(2\alpha_{\bf k})^{-1/2}(a_{\bf k}+a^\dagger_{-{\bf k}})$ and $p_{-{\bf k}}=-i\alpha_{\bf k}(2\alpha_{\bf k})^{-1/2}(a_{\bf k}-a^\dagger_{-{\bf k}})$ with $[a_{\bf k},a^\dagger_{{\bf k}'}]=\delta_{{\bf k},{\bf k}'}$, where $\alpha_{\bf k}=2mE_{\bf k}/(n_sk^2)$ and
\begin{equation}
E_{\bf k}=\sqrt{\frac{2\pi n_se^2}{m}+\left(\frac{k^2}{2m}\right)^2}.
\end{equation}
This finally yields
\begin{equation}
H_h=\sum_{{\bf k}(k<k_0)}E_{\bf k}a^\dagger_{\bf k}a_{\bf k}.
\end{equation}
Equation (9) has the same form as the Bogoliubov spectrum for a weakly coupled charged Bose gas, with the superfluid density $n_s$ taking the place of the condensate density. It will be used in the next Section as the dispersion relation of single-particle excitations in the evaluation of the phase-phase correlations from the single-particle Green's function. Equation (9) should also be contrasted with the dispersion relation of collective excitations, which within a weak-coupling theory is
\begin{equation}
\omega_{\bf k}=\sqrt{\frac{2\pi ne^2}{m}+\left(\frac{k^2}{2m}\right)^2}.
\end{equation}
This relation tends to the 2D plasma frequency $\omega_p=(2\pi ne^2/m)^{1/2}$ at long wavelengths, in accord with the sum rule reported in Eq.~(2). Notice that in the limit $T\rightarrow 0$, where we expect that $n_s\rightarrow n$, 
Eqs.~(9) and (11) give the same result. According to the Gavoret-Nozi\`eres theorem\cite{15}, single-particle and collective excitations coincide at long wavelengths in a Bose-condensed fluid at zero temperature.
	
	We conclude this Section by showing that the sum rule (3) is satisfied in our approach. The response function $\chi_{v_sv_s}({\bf k},\omega)$ for the superfluid velocity is related to the phase-phase response by $\chi_{v_sv_s}({\bf k},\omega)=k^2\chi_{\phi\phi}({\bf k},\omega)/m^2$. At long wavelengths we get from Eq.~(8)
\begin{equation}
\lim_{k\rightarrow0}\chi_{v_sv_s}({\bf k},\omega)=\frac{k^2V_{\bf k}}{2m^2\Omega_p}\left(\frac{1}{\omega-\Omega_p+i\varepsilon}-\frac{1}{\omega+\Omega_p+i\varepsilon}\right)
\end{equation}
where $\varepsilon=0^+$ and $\Omega_p=(2\pi n_se^2/m)^{1/2}$. Insertion of Eq.~(11) into the integral on the LHS of 
Eq.~(3) leads immediately to the desired result.

\section{ASYMPTOTIC BEHAVIOR OF THE SINGLE-PARTICLE GREEN'S FUNCTION}
	We show in this Section that the one-body density matrix of the 2D-BCF at  temperature $T$ below a critical temperature $T_c$ has a power-law decay, as a consequence of the correlations between phase fluctuations in the superfluid. We follow the method proposed for the neutral 2D gas in the work of Popov\cite{5} (see also Fisher and Hohenberg\cite{9}), which derives the power law by evaluating the single-particle Green's function at low momenta from the mean-square fluctuations of the phase $\Phi({\bf r},\tau)$.

	More precisely, the phase fluctuations determine the single-particle Green's function $G({\bf r},\tau;{\bf r}_1,\tau_1)$ in the low-momentum regime (below the cut-off momentum $k_0$) according to
\begin{equation}
G({\bf r},\tau;{\bf r}_1,\tau_1)\propto \exp{\{-\frac{1}{2}\left\langle [\Phi({\bf r},\tau)-\Phi({\bf r}_1,\tau_1)]^2\right\rangle}\}.
\end{equation}
From Eq.~(19.16) in Chapter 6 of Popov's book\cite{5} and using the dispersion relation for single-particle excitations given in Eq.~(9) we find
\begin{equation}
\left\langle \Phi({\bf k},\omega)\Phi(-{\bf k},-\omega)\right\rangle\rightarrow\frac{V_{\bf k}}{\omega^2+n_sV_{\bf k}k^2/m}
\end{equation}
so that
\begin{eqnarray}
\frac{1}{2}\left\langle [\Phi({\bf r},\tau)-\Phi({\bf r}_1,\tau_1)]^2\right\rangle &\rightarrow&\frac{k_BT}{2}\sum_{{\bf k}(k<k_0)}\sum_\omega\frac{V_{\bf k}}{\omega^2+n_sV_{\bf k}k^2/m}\nonumber\\ 
&\times&|\exp{[i({\bf k}\cdot{\bf r}-\omega\tau)]}-\exp{[i({\bf k}\cdot{\bf r}_1-\omega\tau_1)]}|^2.
\end{eqnarray}
The summation over Matsubara frequencies in Eq.~(15) can be carried out explicitly for $\tau_1=\tau^+$, and for $r\equiv |{\bf r}-{\bf r}_1|\rightarrow\infty$ we find
\begin{equation}
\frac{1}{2}\left\langle [\Phi({\bf r},\tau)-\Phi({\bf r}_1,\tau_1)]^2\right\rangle=\frac{r_s}{4}\left(\frac{n}{n_s}\right)^{1/2}\coth\left(\frac{(n_s/n)^{1/2}}{Tr_s}\right)\int_{1/r}^{1/L}\frac{dk}{k}[1-J_0(kr)].
\end{equation}
Here, the temperature $T$ is in units of $e^2/k_B$, $L$ is a length scale of order $1/k_0$, and $J_0(x)$ is the zero-order Bessel function. Finally, the expression for the mean-square phase fluctuation has the form
\begin{equation}
\frac{1}{2}\left\langle [\Phi({\bf r},\tau)-\Phi({\bf r}_1,\tau_1)]^2\right\rangle\rightarrow\alpha\ln(r/L),
\end{equation}
where the quantity $\alpha$ is given by
\begin{equation}
\alpha=\frac{r_s}{4}\left(\frac{n}{n_s}\right)^{1/2}\coth\left(\frac{(n_s/n)^{1/2}}{Tr_s}\right).
\end{equation}
In conclusion, the one-body density matrix of the 2D-BCF decays to zero with the power law $r^{-\alpha}$ as a consequence of the logarithmic correlations between phase fluctuations. In the limit of zero temperature the decay follows the law $r^{-r_s/4}$, which is the result found in the QMC study of Magro and Ceperley\cite{2}.

	Having reached their result from the finite-temperature formula given in Eq.~(18), we can draw two main conclusions: (i) the off-diagonal order reported in the QMC work of Magro and Ceperley\cite{2} proves that the 2D-BCF is a superfluid at $T = 0$; and (ii) a quantum simulation study of the one-body density matrix at finite temperature would provide for the 2D-BCF an alternative approach to the superfluid density, to be compared with the methods that are currently available for its calculation (see {\it e.g.} the method based on the relation between the superfluid density and the mean square winding number\cite{16}). Notice also that, as discussed in the book of Forster\cite{14}, the equivalence between the definition of superfluid density that is provided by the single-particle excitation spectrum and the definition that may be extracted from transverse current correlations and rotation experiments remains a question of continued interest.
	
	As a final point, we should remark that our calculation has not included the possibility of vortex formation in the superfluid as the temperature is raised towards the critical temperature for superfluidity. Equation~(18) should therefore be expected to become invalid near $T_c$.
	
\section{ALTERNATIVE APPROACH TO THE ONE-BODY DENSITY MATRIX}
We present in this Section an alternative calculation of the asymptotic behavior of the one-body density matrix, following an approach which is closely akin to the resummation method adopted in the book of Popov\cite{5} and in the work by Kagan and coworkers\cite{8} for neutral 2D Bose fluids in a quasicondensate state.

	We start from the definition $\rho({\bf r},{\bf r}')=\left\langle\psi^\dagger({\bf r})\psi({\bf r}') \right\rangle$ and use the expression of the field operators in terms of the hydrodynamic density and phase fluctuation operators to find
\begin{equation}
\rho({\bf r},{\bf r}')=n_s-\frac{1}{2}\delta({\bf r}-{\bf r}')+\frac{1}{4n_s}\left\langle\eta({\bf r})\eta({\bf r}') \right\rangle+n_s\left\langle\Phi({\bf r})\Phi({\bf r}') \right\rangle,
\end{equation}
to lowest order in the density and phase fluctuations. The transformations carried out in Section II yield
\begin{equation}		
\left\langle\Phi({\bf r})\Phi({\bf r}') \right\rangle=\sum_{{\bf k}(k<k_0)}\alpha_{\bf k}\left[f(E_{\bf k})+\frac{1}{2}\right]\exp{[i{\bf k}\cdot({\bf r}-{\bf r}')]}
\end{equation}
and a similar expression for $\left\langle\eta({\bf r})\eta({\bf r}') \right\rangle$, with $f(E_{\bf k})=\left\langle a^\dagger_{\bf k}a_{\bf k} \right\rangle=[\exp{(E_{\bf k}/k_BT)}-1]^{-1}$. The final result is
\begin{equation}
\rho(r)=n_s+\sum_{{\bf k}(k<k_0)}\left\{\left(\frac{1}{4n_s\alpha_{\bf k}}+n_s\alpha_{\bf k}\right)\left[f(E_{\bf k})+\frac{1}{2}\right]-\frac{1}{2}\right\}\exp{[i{\bf k}\cdot{\bf r}]}
\end{equation}
where $r=|{\bf r}-{\bf r}'|$. This expression has the same form as in the standard Bogoliubov approach, except that the superfluid density $n_s$ replaces the condensate density $n_0$. Equation~(13) can be re-obtained by an approximate resummation of the phase fluctuations to infinite order.
	
	We write the density matrix in the form\cite{5,8}
\begin{equation}
\rho(r)=\tilde{\rho}(r)\exp{[-\Lambda(r)]}
\end{equation}	
where $\Lambda(r)$ collects all the terms that are responsible for the slow asymptotic decay of $\rho(r)$. We find
\begin{eqnarray}
\Lambda(r)&=&\int\frac{d^2k}{(2\pi)^2}[1-\cos({\bf k}\cdot{\bf r})]\frac{V_{\bf k}}{2E_{\bf k}}\left[2f(E_{\bf k})+1\right]\nonumber\\
&=&\frac{r_s}{4}\left(\frac{n}{n_s}\right)^{1/2}\int_0^\infty dx\frac{1-J_0(xR)}{xg(x)}\left\{1+\frac{2}{\exp{[2(n_s/n)^{1/2}g(x)/(r_sT)]}-1}\right\}
\end{eqnarray}
from Eq.~(21), where $R=2[n_s/(nr_s^2)]^{1/4}(r/l_0)$ and $g(x)=(1+x^4)^{1/2}$. We also find
\begin{eqnarray}
\frac{\tilde{\rho}(r)}{n}&=&1-\frac{1}{n}\int\frac{d^2k}{(2\pi)^2}[1-\cos({\bf k}\cdot{\bf r})]\left\{\frac{k^2}{2mE_{\bf k}}\left[f(E_{\bf k})+\frac{1}{2}\right]-\frac{1}{2}\right\}\nonumber\\
&=&1-r_s\left(\frac{n_s}{n}\right)^{1/2}\int_0^\infty xdx[1-J_0(xR)]\left\{\frac{x^2}{2g(x)}+\frac{x^2}{g(x)(\exp{[2(n_s/n)^{1/2}g(x)/(r_sT)]}-1)}-\frac{1}{2}\right\}.
\end{eqnarray}
In these equations the temperature is scaled in units of $e^2/k_B$.

	A power-law decay of $\rho(r)$ can now be demonstrated analytically from Eqs.~(22)-(24). The function $J_0(xR)$ in the integrand in Eq.~(23) provides a lower limit of integration going as $r^{-1}$, while the upper limit is set by the cut-off wave number $k_0\approx 1/L$ for the applicability of the hydrodynamic Hamiltonian. The asymptotic calculation of the integrals yields
\begin{equation}
\rho(r)\rightarrow \tilde{\rho}_0(r/L)^{-\alpha}
\end{equation}
where the value of the exponent $\alpha$ is given by Eq.~(18) and we have defined
\begin{equation}
\tilde{\rho}_0=\lim_{r\rightarrow\infty}\tilde{\rho}(r).
\end{equation}
In the numerical calculations that are reported in the next Section we find that the ratio $\tilde{\rho}_0/n$ becomes larger than unity at very low temperature, so that this quantity cannot be interpreted as a quasicondensate fraction.

\section{NUMERICAL RESULTS}

In this Section we report some illustrative calculations of the one-body density matrix of the 2D-BCF, based on 
Eqs.~(22) - (24). We need for this purpose to first evaluate the superfluid density as a function of temperature, for which we adopt the Landau theory based on damping of superfluid flow by emission of collective excitations. That is\cite{5,9},
\begin{equation}
\frac{n_s}{n}=1-(2nmk_BT)^{-1}\sum_{{\bf k}\neq 0}k^2\frac{\exp{(\beta\omega_{{\bf k}})}}{[\exp{(\beta\omega_{{\bf k}})}-1]^2}
\end{equation}
with $\beta=1/(k_BT)$ and $\omega_{{\bf k}}$ given by Eq.~(11). This yields
\begin{equation}
\frac{n_s}{n}=1-\frac{2}{T}\int_0^\infty dx\frac{x^3\exp{[2g(x)/(r_sT)]}}{\{\exp{[2g(x)/(r_sT)]}-1\}^2}
\end{equation}
with $T$ in units of $e^2/k_B$. The results for the superfluid fraction $n_s(T)/n$ are shown in Figure 1 for some values of $r_s$ in the weak-coupling regime.

	It is worth noting that, as a consequence of the behavior shown by the superfluid fraction in Figure 1, the single-particle excitation energy $E_{\bf k}$ given in Eq.~(9) starts for $T = 0$ at the collective excitation energy $\omega_{\bf k}$ given in Eq.~(11) and decreases with increasing temperature till its leading long-walength term vanishes at the critical temperature $T_c$. The plasmon gap in the collective excitation spectrum of Eq.~(11) remains instead constant with increasing temperature. This behavior of the plasmon gap in the single-particle spectrum of a charged boson fluid has previously been found in the 3D case within the Hartree-Fock-Bogoliubov approximation\cite{17}, with small deviations from the predictions of the Gavoret-Nozi\`{e}res theorem being present at $T = 0$.
	
	Our results for the one-body density matrix are reported in Figures 2 and 3, first at $T = 0$ for several values of $r_s$ (Figure 2) and then at $r_s=0.1$ for several values of $T$ (Figure 3). In each of these Figures the right-hand panel shows a logarithmic plot of the numerical results for $\rho(r)/n$ and a linear fit based on Eq.~(25), using the value of the exponent $\alpha$ given by Eq.~(18). The length $L$ in Eq.~(25), which is not determined by the asymptotic calculation, appears in a logarithmic scale as an additive constant and has been obtained from the linear fitting procedure at large $r$. Evidently the power-law prediction is in excellent agreement with the numerical results over an extended range of values for the reduced distance .
	
\section{SUMMARY}

	In summary, we have studied superfluidity in a weakly interacting 2D fluid of charged bosons with $\ln(r)$ interactions as a function of temperature, using a hydrodynamic reduction of the Hamiltonian that has allowed us to treat the system as a superfluid in the absence of a Bose-Einstein condensate at zero temperature. In the Appendix we show that the absence of a condensate at $T = 0$ also follows for this system from the standard Bogoliubov approach.
	
	We have shown that the assumption of complete superfluidity for this system at $T = 0$ accounts quantitatively for the QMC data of Magro and Ceperley\cite{2} on the power-law decay of the one-body density matrix. In turn, computer studies at finite temperature would allow one to determine the behavior of the superfluid fraction and provide a full quantitative test of our theory.

\acknowledgements
{
	This work was partially supported by MIUR under the PRIN-2000 Initiative.
}

\renewcommand{\theequation}{A.\arabic{equation}}
\setcounter{equation}{0}
\section*{APPENDIX. ABSENCE OF BOSE-EINSTEIN CONDENSATION IN THE 2D-BCF}

	We show here that within the Bogoliubov approximation\cite{12} there is no Bose-Einstein condensate in the 2D-BCF both at zero and at finite temperature. In contrast, in the 2D fluid of charged bosons with $e^2/r$ interactions a condensate is present at $T = 0$ and the notion of quasicondensation can be used at $T\neq 0$\cite{11}.
	
	If we assume that the condensate fraction $n_0/n$ is non-zero, then it is easy to show that the expression (9) for the single-particle excitation energy becomes
\begin{equation}
E_{\bf k}=\sqrt{\frac{2\pi n_0e^2}{m}+\left(\frac{k^2}{2m}\right)^2}
\end{equation}
in the Bogoliubov approach. The corresponding equation for the condensate fraction reads
\begin{equation}
\left(1-\frac{n_0}{n}\right)\left(\frac{n_0}{n}\right)^{-1/2}=\frac{r_s}{2}\int_0^\infty xdx\left\{\frac{h(x)}{2x^2g(x)}-1+\frac{[x^2g(x)]^{-1}}{\exp{[2(n_0/n)^{1/2}g(x)/(r_sT)]}-1}\right\}
\end{equation}
where $h(x)=1+2x^4$.

	The first term in the square brackets on the RHS of Eq.~(A.2) gives a contribution of order $x^{-1}$ to the integrand as $x\rightarrow0$, and hence the integral diverges. We conclude that $n_0/n$ must vanish at any temperature, in agreement with the general sum-rule argument given by Magro and Ceperley\cite{2}.
	The property $n_0/n=0$ is consistent, on the other hand, with the Bogoliubov - des Gennes equations for the Bogoliubov amplitudes $u_{\bf k}$ and $v_{\bf k}$. In this case these equations yield $v_{\bf k}= 0$ and $u_{\bf k}=1$, so that the particle density is related to the (now finite) chemical potential by
\begin{equation}
n=\sum_{{\bf k}\neq 0}\left\{\exp{\left[\beta\left(\frac{k^2}{2m}-\mu\right)\right]}-1\right\}^{-1}.
\end{equation}

\newpage 

\begin{figure}
\centerline{\mbox{\psfig{figure=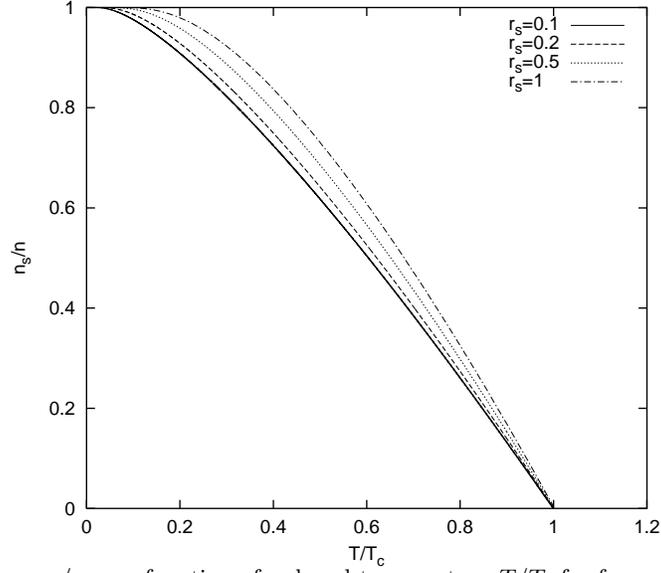, angle =0, width =9 cm}}} 
\caption{The superfluid fraction $n_s/n$ as a function of reduced temperature $T/T_c$ for four values of the coupling strength $r_s$. In absolute units the critical temperature $T_c$ takes the values 136.9, 41.3, 8.9 and 2.9 $e^2/k_B$ for $r_s$ going from 0.1 to 1.}
\label{Fig1}
\end{figure}

\begin{figure}
\centerline{\mbox{\psfig{figure=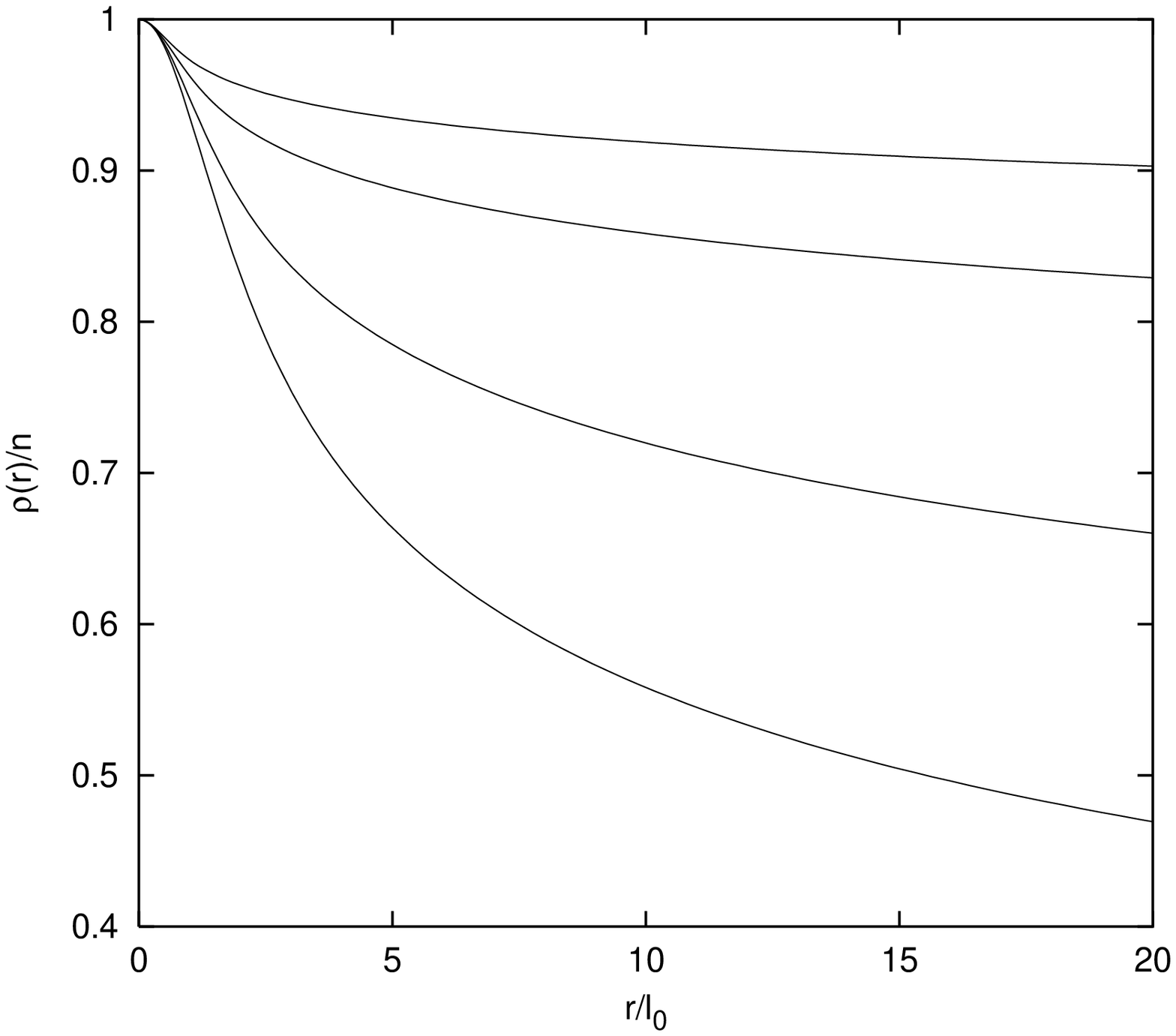, angle =0, width =8 cm}\psfig{figure=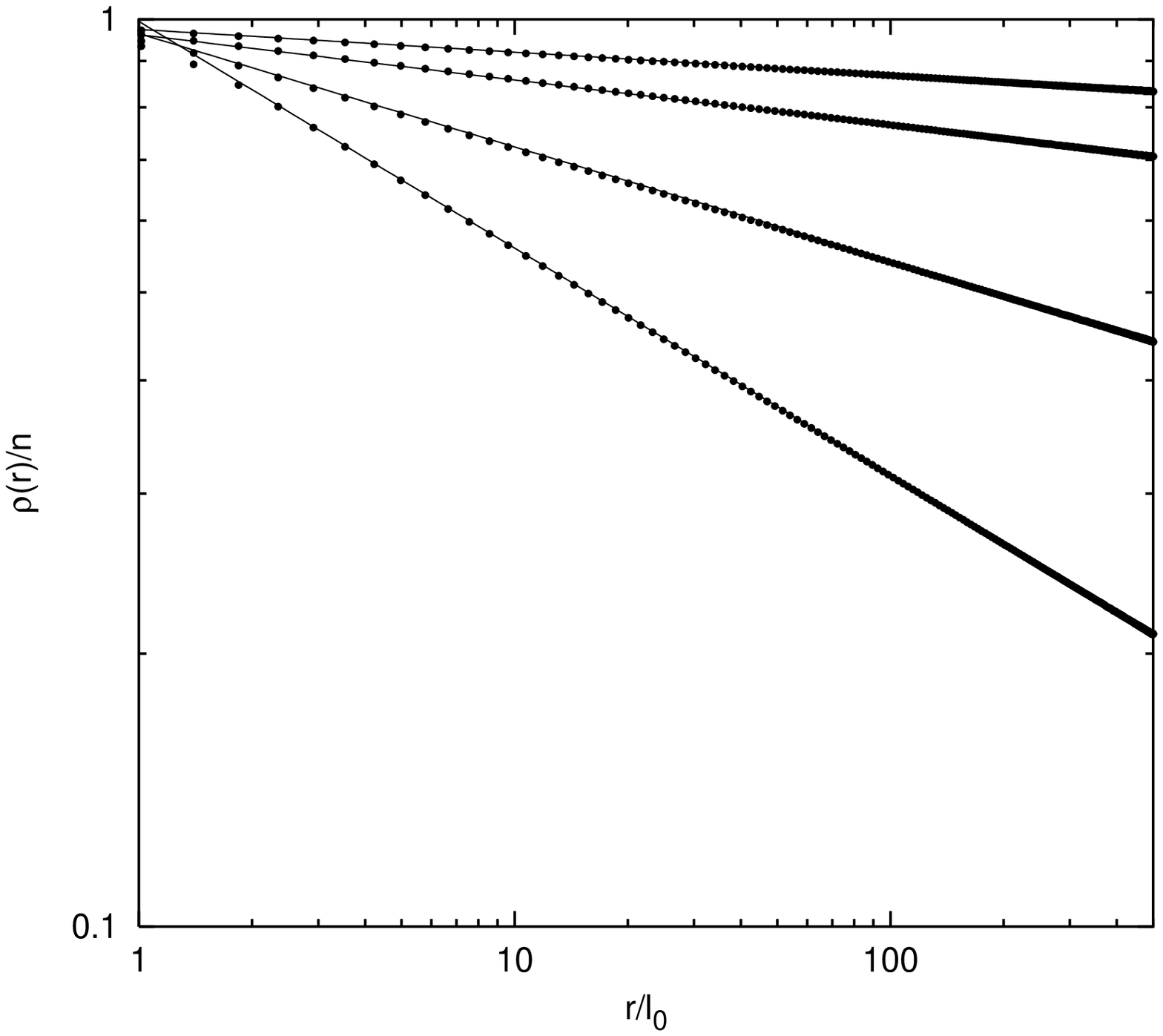, angle =0, width =8 cm}}} 
\caption{The one-body density matrix $\rho(r)/n$ as a function of distance $r$ (in units of the reference length $l_0$) in the 2D-BCF. Left: at $T = 0$ for $r_s = 0.1,\; 0.2,\; 0.5,\; {\rm and}\; 1$ (from top to bottom). Right: on a logarithmic scale, the same numerical results are shown as dots while the predictions from the Green's function approach are shown as continuous lines.}
\label{Fig2}
\end{figure}

\begin{figure}
\centerline{\mbox{\psfig{figure=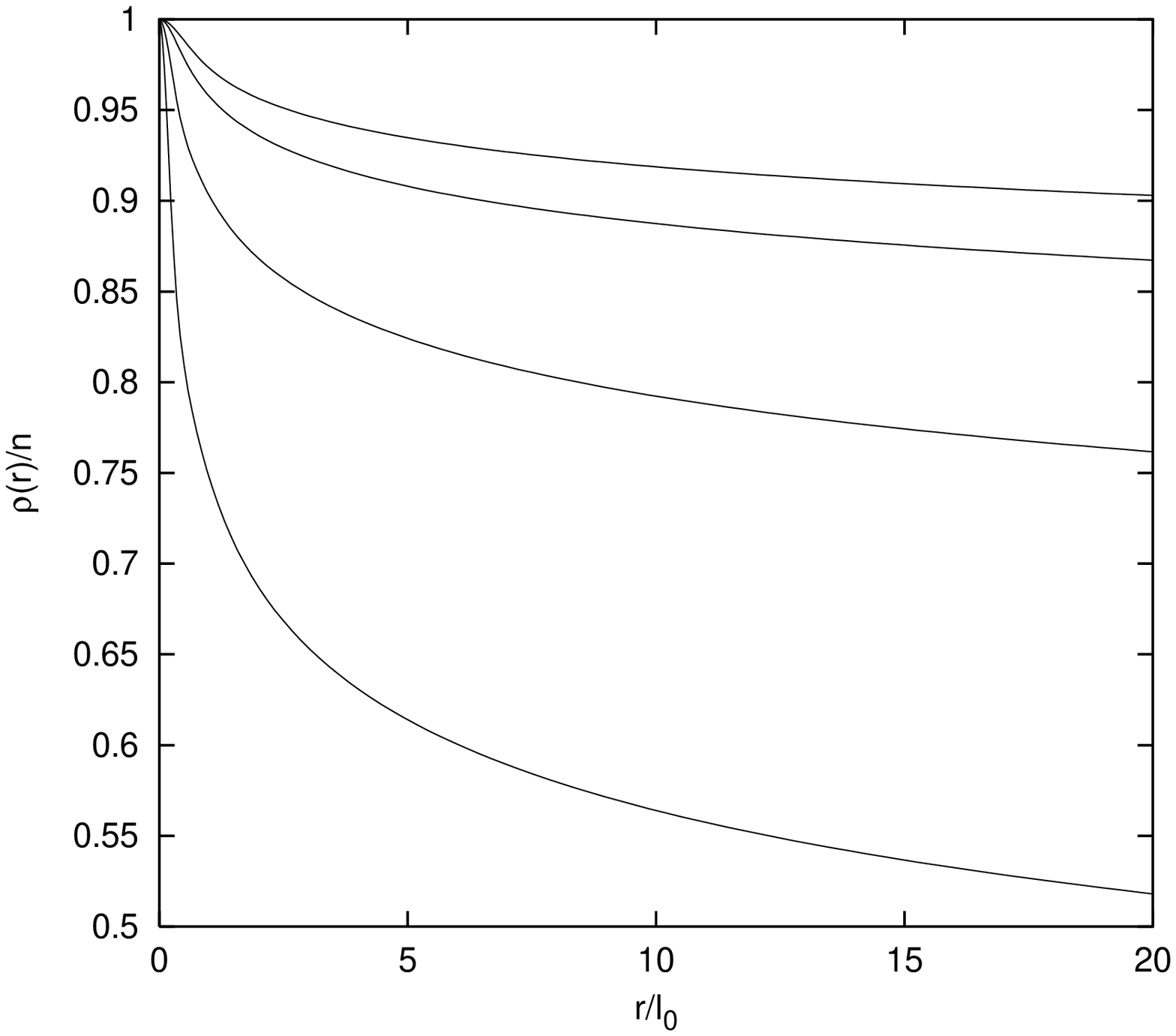, angle =0, width =8 cm}\psfig{figure=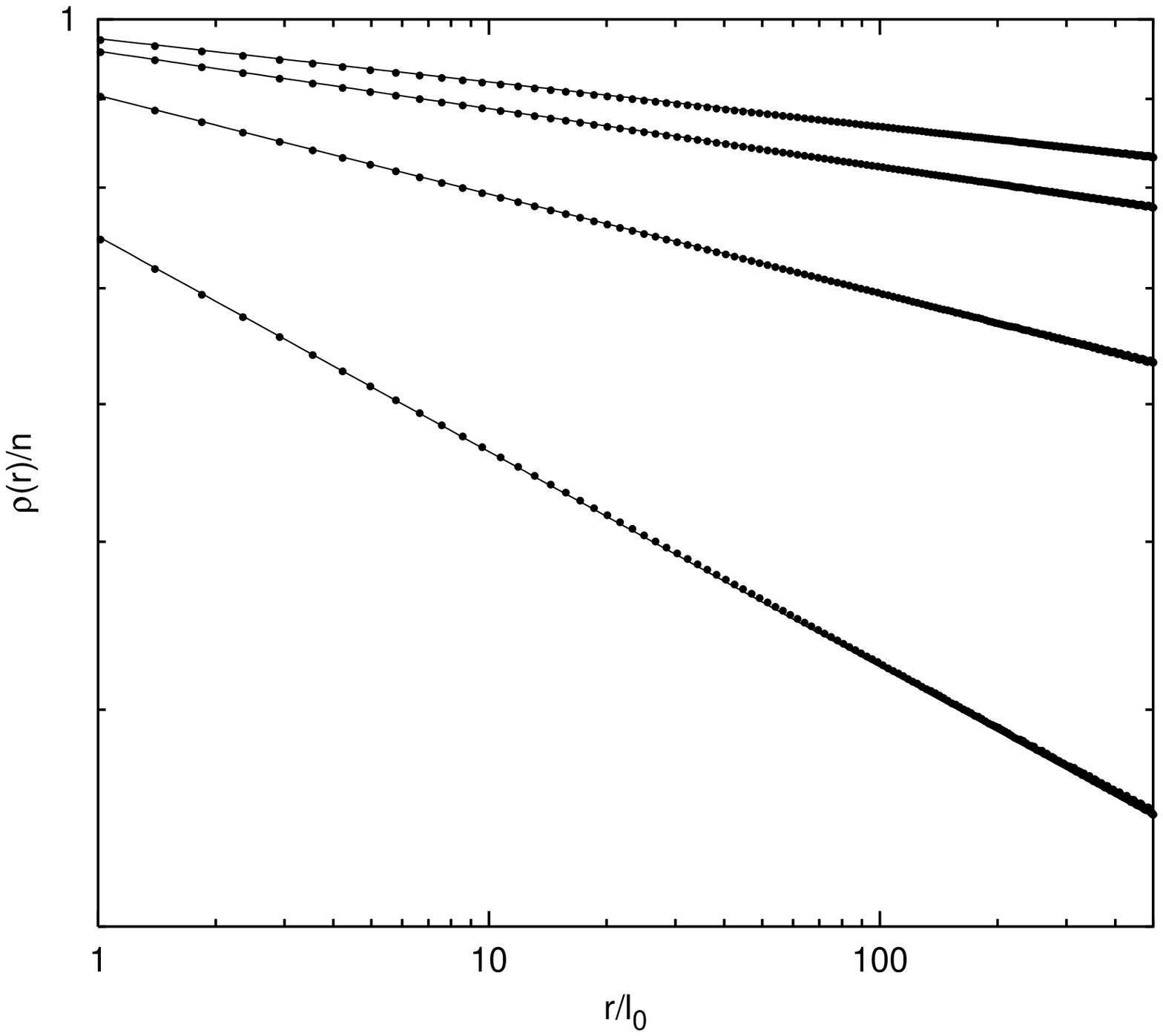, angle =0, width =8 cm}}}
\caption{The one-body density matrix $\rho(r)/n$ as a function of distance $r$ (in units of $l_0$) in the 2D-BCF. Left: at $r_s = 0.1$ for $T = 0,\; 10,\; 20,\; {\rm and}\; 40$ $e^2/k_B$ (from top to bottom). Right: on a logarithmic scale, the same numerical results are shown as dots while the predictions from the Green's function approach are shown as continuous lines.}
\label{Fig3}
\end{figure}
\end{document}